%% file: main.tex
\documentclass[11pt]{llncs}
\usepackage{latexsym}

\begin{document}

\title{Active Supervisory Control for Opacity Preservation
}
\author{Damas P. Gruska }

\institute{
Department of Applied Informatics, Comenius University,\\
Mlynska dolina, 842 48 Bratislava, Slovakia,\\
gruska@fmph.uniba.sk.}

\pagestyle{empty}

\maketitle

\begin{abstract}

Algebraic methods are employed to define language-based security properties of processes. An active supervisor is introduced that can disable unwanted behavior of an insecure process by controlling some of its actions or by inserting timed actions to make the process secure. We assume a scenario where neither the supervisor nor the attacker has complete information about the ongoing system's behavior. We study the conditions under which such a supervisor exists, as well as its properties and limitations. We present a series of decidability and undecidability results for various types of systems, observations, attackers, and supervisors. Finally, we focus on timing attacks.

\end{abstract}

\input{Intro}

\input{formalism}

\input{opacity}
\input{supervisor}

\input{Conclusion}

\end{document}

%% file: Intro.tex
\section{Introduction}

The deployment of formal models and rigorous verification methods to analyze, define, and discover system vulnerabilities regarding security threats has become a paramount area of research in modern systems engineering. Formal methods provide a mathematically sound framework that can, in many instances, definitively prove whether a given system is susceptible to security breaches. When a system is found to be insecure, designers generally face two primary courses of action. The first option is to fundamentally re-design the system's behavior. However, this approach is often prohibitively costly, complex, or entirely infeasible—particularly when the system is already embedded in fixed hardware architectures, proprietary firmware, or legacy software ecosystems.

The second, more flexible option is to employ supervisory control theories (see e.g., \cite{W70,RW89,RRF21,KC24,XTWS25}) or related architectural mitigations (\cite{HAM15}) to dynamically restrict the system's execution paths. Under traditional supervisory control frameworks, an external supervisor monitors a subset of observable system events and selectively enables or disables controllable actions. By dynamically filtering these actions, the supervisor forces the underlying system to adhere to a specified safety or security policy (see also \cite{GR18}). Naturally, this enforcement introduces a fundamental trade-off between strict security guarantees and optimal system functionality, as over-restricting behavior can render a system non-operational or inefficient.

However, this classical paradigm of restricting actions falls short when dealing with timing attacks. As a potent class of side-channel attacks, timing attacks exploit timing information leakage—a phenomenon where an adversary infers internal, highly sensitive (private) data simply by measuring variations in execution and response times. Because they bypass traditional logical boundaries, timing attacks pose a critical threat, enabling intruders to compromise theoretically "unbreakable" cryptographic systems, algorithms, and cryptographic protocols. In such scenarios, merely blocking or permitting actions is insufficient, as the mere pacing of the events themselves leaks critical data.

To mitigate these side-channel vulnerabilities, systems must look beyond logical enforcement and manipulate temporal behavior. One established countermeasure is the strategic insertion of artificial time delays between individual actions, effectively smoothing out temporal variances and blinding the attacker to internal state changes. Bridging these two paradigms, this paper introduces a novel framework for an active supervisor. Distinct from traditional models, our proposed supervisor possesses a dual capability: it not only retains the authority to restrict unauthorized actions but can also dynamically inject calculated time delays into the execution stream, thereby robustly guaranteeing the language-based opacity and safety of the underlying process.

Regarding the underlying formalism, this framework builds upon a language-based variant of a security property known as opacity (see \cite{BKMR06,BKR04}), integrated within a timed process algebra framework. While traditional security models often rely on finite automata or Timed Automata, process algebras offer a distinct advantage: they provide a rich set of compositional constructs. This compositionality allows us to naturally and modularly express not only the system behaviors and observations, but also the intruders and active supervisors themselves as distinct, interacting algebraic processes. Furthermore, this timed algebraic setting provides the precise mathematical machinery required to formally model and reason about timing attacks and information leakage.

This work directly extends our previous research. In \cite{Gru19,Gru21a,Gru26}, we investigated the conditions under which a timed insertion function exists to enforce state-based security properties for a given process, establishing a baseline of decidability and undecidability results. However, state-based models can be limiting when security policies depend on the entire history of actions rather than static system states. To address this, the current study shifts the focus to language-based opacity, offering a more expressive framework for tracking execution paths. Crucially, we also move beyond purely passive mitigation by combining these timed insertion functions with active supervisors, creating a unified control mechanism capable of both blocking unauthorized actions and dynamically manipulating event timings to guarantee process safety.

Supervisors that can guarantee the security of systems concerning opacity are often studied for finite automata or other models with a finite number of states. Here we work with a formalism equivalent to a Turing machine (see \cite{Mil89}), which introduces new challenges. We show whether the existence of a supervisor is decidable or undecidable. In addition, we assume more general observations, not only those that hide some actions.

\noindent {\em Main contributions of this paper:} 
\begin{itemize}
    \item \textbf{Novel Active Supervisor Framework under Partial Visibility:} We introduce an active supervisor model operating under asymmetric and incomplete information regarding system execution. Unlike traditional supervisors that only restrict behavior, this model combines action-disabling capabilities with the dynamic insertion of timed actions to actively mitigate information leakage.
    \item \textbf{Asymmetric Observation Functions:} We formalize and analyze scenarios where the attacker and the supervisor possess different observation functions. This allows us to model realistic, non-trivial security settings where neither entity has complete visibility, and their views of the system's execution history diverge.
    \item \textbf{Delineation of Decidability Boundaries:} We establish a comprehensive set of decidability and undecidability results concerning the existence of such supervisors. We chart these boundaries across various system types, observation granularities, and capabilities of both attackers and supervisors, mapping out exactly where verification remains feasible.
\end{itemize}

The remainder of this paper is organized as follows. Section 2 outlines the foundational preliminaries, introducing the syntax and operational semantics of the Timed Process Algebra (TPA) that serves as our underlying formalism. In Section 3, we formalize the notion of language-based security properties, focusing specifically on opacity within this timed framework. Section 4 introduces our core contribution: the formal definition of the active supervisor, alongside an analysis of its behavior, properties, and structural limitations under partial observability. Here our theoretical results, establishing the precise boundaries of decidability and undecidability for various system and attacker configurations are presented. Finally, Section 5 concludes the paper with a summary of our findings and a discussion of promising directions for future research.

%% file: formalism.tex
\section{Working Formalism}

In this section, we briefly recall  
our working formalism which will be based on 
Timed Process Algebra, TPA for short.
TPA itself
is based on Milner's Calculus of Communicating Systems (for short, CCS, 
see \cite{Mil89}), so we will start with this. 
To define the language CCS, we first presuppose a set of atomic action
symbols $A$ not containing symbols $\tau$, and that for every $a \in A$
 there exists an $\bar a \in A$ and $\bar{\bar a} =a$ i.e.  actions which represent receiving and sending from a channel $a$, respectively.
We define $Act = A \cup \{ \tau\}$ where $\tau$ represents internal action, for example, a result of internal communication. 
We let $a,b,\dots $ range over $A$; $x, y,\dots $ range over $Act$.
 The set of CSS terms is defined by the following
BNF expression where  $x \in Act, X \in Var$, $Var$ is a set of process
variables,
 $f$ is ranging over
relabelling functions, $f:Act \to Act$ is such that
$\overline{f(a)}=f(\bar a) $ for $a \in A, f(\tau)=\tau$ and
finally,
$L \subseteq A$:

$$ P::= Nil \, \,\vert \, \,  X  \, \,  \mid  \, \,  x.P  \, \,  
\mid  \, \,  P+P  \, \,  \mid  \, \,  P \mid P   \, \, \mid  \, 
\, \mu XP \, \,
   \mid \, \,  P \setminus L  \, \,  \mid  \, \,  P[f] $$

We use the usual definitions for free and bound variables,
open and closed terms and guarded recursion.
The set {\bf CCS} of {\it processes}  consists of closed and   
guarded CCS terms.

$Nil$ represents process doing nothing,  $X$ is a process variable.  
Process $  x.P $ can perform action $x$ and then behaves as process $P$ 
(prefix operator) ,  $+,  \mid$ represent nondeterministic choice and parallel operator,  respectively. 
A recursion operator is denoted by $\mu XP$ i.e. recursive definition of the process
given by equation $X=P$.  The unary operators $P \setminus L$ and $P[f] $ represent
restriction and relabeling, respectively. 
Formal definition of a structural operational semantics for {\bf CCS}  is defined in terms of
Labeled 
Transition Systems.

\begin{definition} \rm  {\it A Labeled Transition System} is a triple
$(S,T,\{\stackrel{x}{\to},x \in T\})$ where $S$ is a set of states, 
$T$ is a set of labels and $\{\stackrel{x}{\to},x \in T\}$ is the
transition
relation such that each $\stackrel{x}{\to}$ is a binary transition relation
on $S$.
We write $P \stackrel{x}{\to} P'$ instead of
$(P,P') \in \stackrel{x}{\to} $ and  $P  \stackrel{x}{ \not \to} $ if
there is no $P'$ such that $P \stackrel{x}{\to} P'$.
\end{definition}

As a set of states we use the set of CCS terms, the set of labels is equal to $Act$, and transition relations are defined as follows.

\begin{definition} The  transition relations 
$T=\{\stackrel{x}{\to}_{ CCS},x \in Act\}$
are defined as the least relations satisfying the following 
inference rules:

$$
\begin{array}{cc} 
\displaystyle 
\frac {\: }
{\: x.P \stackrel{x}{\rightarrow}  P   \:}
&
\displaystyle   
\frac {\: P \stackrel{x}{\rightarrow} P' \:  }
{\: P + Q \stackrel{x}{\rightarrow} P',
 \: Q + P \stackrel{x}{\rightarrow} P'\:} 
\\
 & 
\\
\displaystyle 
\frac {\: P \stackrel{x}{\rightarrow} P' \:  }
{\: P \setminus L \stackrel{x}{\rightarrow}  P'\setminus L \:},
(x,\bar x \not \in L)
&
\displaystyle 
 \frac {\: P[\mu XP/X] \stackrel{x}{\rightarrow} P' \: }
            {\: \mu XP \stackrel{x}{\rightarrow} P' \: } 
\\
 & 
\\
\displaystyle
\frac {\: P \stackrel{x}{\rightarrow} P' }
			{\: P[f] \stackrel{f(x)}{\rightarrow} P'[f] \:}
&
\displaystyle
\frac {\: P \stackrel{a}{\rightarrow} P' ,  
Q \stackrel{\bar a}{\rightarrow}Q' \:}
{\: P \mid Q  \stackrel{\tau}{\rightarrow} P' \mid Q' \:}
\\
&
\\
\displaystyle
\frac {\: P \stackrel{u}{\rightarrow} P'  \:}
{\: P \mid Q  \stackrel{u}{\rightarrow} P' \mid Q,
Q \mid P  \stackrel{u}{\rightarrow} Q \mid P' \:}
 & 
\end{array}
$$

\end{definition}

We are now ready to define the working formalism, which is the time extension of CCS.
In TPA we use  the special time action $t$ which expresses elapsing of
(discrete) time is added and hence the set of actions is extended from $Act$ to $Actt$. 
The presented language is a slight
simplification of  Timed Security Process Algebra (tSPA) introduced in
\cite{fabio}. We omit an explicit idling operator $\iota$ used in
tSPA and instead of this we allow implicit idling of processes.
Hence processes can perform either "enforced idling" by performing
$t$ actions which are explicitly expressed in their descriptions or
"voluntary idling" (i.e. for example, the process $a.Nil$ can perform $t$ action
despite the fact that this action is not formally expressed in the process specification). But in 
both cases, internal communications have priority to action $t$ in
the parallel composition. Moreover, we do not divide actions into
private and public ones as it is in tSPA. TPA differs also from the
tCryptoSPA (see \cite{gor04}). TPA does not use value passing and
strictly preserves {\em time determinancy} in case of choice operator
$+$ what is not the case of tCryptoSPA (see \cite{Gru15}).
To define  $At
= A \cup \{ t \}, Actt = Act \cup \{ t \}$, moreover we suppose that $S(t)=t$ for every relabelling function $S$. 
We give a structural operational semantics of terms again by means of
labeled transition systems. 
 The set of terms represents a set of
states, labels are actions from $Actt$. The transition relation
$\to$ is a subset of $\mbox{TPA} \times Actt
 \times \mbox{TPA} $.  
We
define the transition relation as the least relation satisfying the
inference rules for CCS plus the following inference rules for $t$ action (for more details see \cite{Gru15}).:

$$
\begin{array}{clccl}
\displaystyle \frac {\:  \: }
      {\: Nil  \stackrel{t}{\rightarrow} Nil  \:}
& A 1 & \mbox{} & \displaystyle \frac {\:  \: }
      {\: u.P  \stackrel{t}{\rightarrow} u.P  \:}
& A 2
\\
&
\\
\displaystyle \frac {\:  P \stackrel{t}{\rightarrow} P' , Q
\stackrel{t}{\rightarrow}  Q', P \mid Q 
\stackrel{\tau}{\not \to}  \:} {\: P \mid Q
\stackrel{t}{\rightarrow} P' \mid Q' \:} &Pa
 &
\mbox{ } & \displaystyle \frac {\: P \stackrel{t}{\rightarrow} P', Q
\stackrel{t}{\rightarrow} Q' \:  } {\: P + Q
\stackrel{t}{\rightarrow} P' +  Q'\:} & S
\end{array}
$$

For $s = x_1.x_2. \dots .x_n, x_i \in Act$ we write $P
\stackrel{s}{\rightarrow}$ instead of $P
\stackrel{x_1}{\rightarrow}\stackrel{x_2}{\rightarrow} \dots
\stackrel{x_n}{\rightarrow}$ and we say that  $s$ is a trace of $P$.
The set of all traces of $P$ will be denoted by $Tr(P)$ i.e.  $ Tr(P) = \{ s | P \stackrel{s}{\rightarrow}\}$. By
$\epsilon$ we denote the empty sequence and by $M^*$ we denote the set of finite sequences of elements from $M$.
Let $N \subseteq M$ and $w \in M^*$. By $s|N$ we denote the sequence obtained from $s$ by removing all actions belonging to $N$. 
Let $s,s' \in M^*$ by $s \sqsubseteq s'$ we denote that $s$ is a prefix of $s'$.
We use $\stackrel{x}{\Rightarrow} $  as an abbreviation for transitions including $\tau$ actions
 i.e. 
$(\stackrel{\tau}{\rightarrow})^*\stackrel{x}{\rightarrow}(\stackrel{\tau}{\rightarrow})^* $
  (see \cite{Mil89}).
By $\lambda(P)$ we will denote a set of succesors of $P$ i.e. 
 $\lambda(P) = \{ P' | P \stackrel{s}{\rightarrow} P', s \in Actt^* \}$.



Now we define two behavior  equivalences,  namely weak trace equivalence and bisimulation, respectively (see \cite{Mil89}).

\begin{definition}
The set of  weak traces of  process $P$  
 is defined as
$Tr_{w}(P) = \{s \in (A \cup \{t\})^\star | \exists P'. P
\stackrel{s}{\Rightarrow} P' \}$. 
Two processes $P$ and $Q$ are  weak trace equivalent (denoted as $P \approx_w Q$) iff $Tr_{w}(P) = Tr_{w}(Q)$.
Two processes $P$ and $Q$ are   trace equivalent (denoted as $P \approx Q$) iff $Tr(P) = Tr(Q)$.

\end{definition}

\begin{definition} \label{simulation} Let $(\mbox{TPA},Act, \to)$ be a labeled
transition system (LTS). A relation $\Re \subseteq \mbox{TPA} \times
\mbox{TPA}$ is called a {\em  bisimulation} if it is
symmetric and it satisfies the following condition: if $(P, Q) \in
\Re $ and $P \stackrel{x}{\rightarrow}  P', x \in Actt$ then there
exists a process $Q'$ such that $Q
\stackrel{x}{\rightarrow} Q'$ and $(P', Q') \in \Re $.
Two processes $P,Q$ are {\it  bisimilar}, abbreviated $P \sim
Q$, if there exists a   bisimulation relating $P$ and $Q$.
If it is not required that relation $\Re$ is symmetric we call it
simulation and we say that process $P$ simulates process $Q$,
abbreviated $P \prec_s Q$, if there exists a simulation relating $P$
and $Q$.
\end{definition}

%% file: opacity.tex
\section{Opacity}

System security based on the absence of information flow relies on the assumption that an adversary observing the system's execution is unable to deduce its classified or secret properties. This fundamental paradigm of information flow security was pioneered by Goguen and Meseguer in their seminal work on noninterference \cite{GM82}. To formalize and reason about information flow, traditional approaches often partition the system's action alphabet into strictly public and private sets at the description level, as seen for instance in \cite{gor04}. In contrast, we adopt a more general and flexible framework rooted in the concepts of observation functions and opacity. The concept of opacity was originally introduced and exploited in \cite{BKR04} and \cite{BKMR06} within the context of Petri nets and transition systems, respectively, to characterize whether a system's secret remains hidden from an outside observer.

To formalize this in our setting, we first define an observation function ${\cal  O}$ over execution sequences from $Actt^\star$. Rather than assuming a static, uniform view of system actions, we recognize that an attacker's or supervisor's visibility may be dynamic. Consequently, various configurations of observation functions can be formulated depending on the structural and environmental contexts they take into account. For example, a context-dependent observation function might dictate that the visibility or interpretation of a specific action depends inherently on the history of preceding actions, the current clock time, or local state changes. By varying the capabilities of these observation functions, we can model a wide spectrum of adversarial scenarios—ranging from completely oblivious onlookers to highly sophisticated, context-aware intruders.

\begin{definition} \label{obs} {\bf (Observation)} 
Let $\Theta$ be a set of elements called observables. Any function
${\cal  O}: Actt^\star \rightarrow \Theta^\star$ is an observation
function. It is called static /dynamic /orwellian / m-orwellian $(m
\geq 1)$ if the following conditions hold respectively (below we
assume $w=x_1 \dots x_n$):

\begin{itemize}
\item static if there is a mapping
${\cal  O}': Actt \rightarrow \Theta \cup \{\epsilon\}$ such that
for every $w \in Actt^\star$ it holds ${\cal  O}(w) = {\cal O}'(x_1)
\dots {\cal  O}'(x_n)$,

\item dynamic if
there is a mapping ${\cal  O}': Actt^\star \rightarrow \Theta \cup
\{\epsilon\}$ such that for every $w \in  Actt^\star$ it holds
${\cal  O}(w) = {\cal  O}'(x_1).{\cal  O}'(x_1.x_2) \dots {\cal
O}'(x_1 \dots x_n)$,

\item orwellian if
there is a mapping ${\cal  O}': Actt \times Actt^\star \rightarrow
\Theta \cup \{\epsilon\}$ such that for every $w \in Actt^\star$ it
holds
 ${\cal  O}(w) = {\cal  O}'(x_1,w).{\cal  O}'(x_2,w) \dots {\cal  O}'(x_n,w)$,

\item m-orwellian if
there is a mapping ${\cal  O}': Actt \times Actt^\star \rightarrow
\Theta \cup \{\epsilon\}$ such that for every $w \in Actt^\star$ it
holds
${\cal  O}(w) = {\cal  O}'(x_1,w_1).{\cal  O}'(x_2,w_2) \dots {\cal
O}'(x_n,w_n)$ where
$w_i = x_{max\{1, i-m+1\}}.x_{max\{1, i-m+1\}+1} \dots x_{min\{n,
i+m-1\}}$.
\end{itemize}

\end{definition}

In the case of the static observation function, each action is
observed independently from its context. In the case of the dynamic
observation function, an observation of an action depends on the
previous ones, in the case of the orwellian and m-orwellian
observation function an observation of an action depends on the all
and on $m$ previous actions in the sequence, respectively. The
static observation function is the special case of m-orwellian one
for $m=1$. Note that from the practical point of view the
m-orwellian observation functions are the most interesting ones. An
observation expresses what an observer - eavesdropper can see from a
system behavior and we will alternatively use both the terms
(observation - observer) with the same meaning.
Note that the same action can be seen differently during an observation (except static observation function) and this expresses a possibility of accumulating some knowledge by an intruder. For example, an action not visible at the beginning could become somehow 
observable. An observation function can be naturally extended to a set of sequences.  
 From now on, we assume that $\Theta \subseteq Actt$. We now define observation functions that are independent of any actions from a set $M$ appearing in the input.

\begin{definition} \label{inde}
A function ${\cal O}: Actt^\star \rightarrow \Theta^\star$ is called 
$M$-independent for $M \subseteq Actt$ if and only if 
${\cal O}(w) = {\cal O}(w | M)$ for every $w \in Actt^*$. We say that ${\cal O}$ is 
monotonic if $w \sqsubseteq w'$ implies ${\cal O}(w) \sqsubseteq {\cal O}(w')$.
\end{definition}

From now on, we assume only observation functions which are (1) $\{\tau\}$-independent, (2) not constant, and (3)  monotonic. The first requirement states that observations do not depend on the performance of the internal action $\tau$, and the second implies that observations depend on the inputs, and the third that longer inputs produce longer observations.
Now, suppose there is a security property over process traces expressed by a predicate $\phi$. Such a property could represent the execution of one or more classified actions, or the execution of actions in a specific confidential order. In general, we assume that this predicate holds for some traces and not for others. We aim to determine whether an observer can deduce the validity of the property $\phi$ solely by observing the action sequences from $Actt^\star$ performed by a given process.
The observer cannot deduce the validity of $\phi$ if there are two
traces $w,w' \in Actt^\star$ such that $\phi(w), \neg \phi(w')$ and
the traces cannot be distinguished by the observer, i.e. ${\mathcal O}(w)
= {\mathcal  O}(w')$. We formalize this concept by opacity.

\begin{definition} \label{opacity} {\bf (Language Opacity) }
Given process $P$, a predicate $\phi$ over $Actt^\star$ is language opaque w.r.t. the observation function ${\mathcal  O}$ if for every sequence
$w$, $w \in Tr(P)$ such that $\phi(w)$ holds and ${\mathcal  O}(w) \not
= \epsilon$, there exists a sequence
 $w', w' \in Tr(P)$ such that $\neg \phi(w') $ holds
 and ${\mathcal  O}(w) = {\mathcal  O}(w')$.
 The set of processes for which the predicate $\phi$
is language opaque with respect to ${\mathcal  O}$ will be denoted by
$LOp^{\phi}_{\mathcal O}$.
\end{definition}

 A predicate is opaque if, for any system trace for which the predicate holds, there exists another system trace for which it does not hold, such that both traces are indistinguishable to an observer (as defined by an observation function). Consequently, the observer (intruder) cannot definitively deduce whether the executed trace satisfies the predicate.

Suppose that the set of all actions $A$ is partitioned into two disjoint sets: 
public (low-level) actions $L$ and private (high-level) actions $H$, such that 
$A = L \cup H$ and $L \cap H = \emptyset$. 
The \emph{Strong Nondeterministic Non-Interference} (SNNI, see \cite{fabio}) 
property assumes an intruder who attempts to deduce whether any private action 
has been performed by a given process, while being able to observe only public 
actions. If the intruder cannot infer this information, the process satisfies 
the SNNI property. 
Note that the SNNI property is a special case of opacity. This holds under a 
static observation function $\mathcal{O}: A^* \to L^*$, homomorphically 
extended from actions to sequences, defined for an action $a \in A$ as:
${\mathcal  O}(x)=x$ iff $x \not \in H$ and 
 ${\mathcal  O}(x)=\epsilon$ otherwise,
and a predicate $\phi(w)$ that evaluates to true if and only if the trace $w$ 
contains at least one action from $H$.

 Language opacity expresses the security of processes but says nothing about the security of their successors. Consequently, a system might be language opaque initially, yet transition into a state that completely compromises its secrets. To address this limitation, we define a stronger property called persistent language opacity, which ensures that security guarantees are maintained invariant under process execution.

\begin{definition} \label{def:persistent_opacity} {\bf (Persistent Language Opacity)}
Given a process $P$, a predicate $\phi$ over $Actt^\star$ is persistently language opaque 
w.r.t.\ the observation function ${\mathcal O}$ if 
for every $P' \in \lambda(P)$, we have $P' \in LOp^{\phi}_{\mathcal O}$.
The set of processes for which the predicate $\phi$ 
is persistently language opaque with respect to ${\mathcal O}$ is denoted by 
$PLOp^{\phi}_{\mathcal O}$.
\end{definition}
 
Clearly, we have $PLOp^{\phi}_{\mathcal{O}} \subseteq LOp^{\phi}_{\mathcal{O}}$ since $P \in \lambda(P)$. However, the following example demonstrates that persistent language opacity is a stronger property than language opacity.

\begin{example}
Let $\phi(w)$ holds iff $w$ contains action $h$ and ${\mathcal  O}$ is a static observation function such that  ${\mathcal  O}(h) = \epsilon$ and ${\mathcal  O}(x) = x$ for $x \not = h$. 
Then for $P_1= l.h.l.Nil$, $P_2 = l.h.l.Nil + l.l.Nil$ we have 
$P_1 \not \in LOp^{\phi}_{\mathcal O}$,
$P_1 \in LOp^{\phi}_{\mathcal O}$
and
$P_2 \not \in PLOp^{\phi}_{\mathcal O}$.
\end{example}

We can define a partial ordering on the set of observation functions that reflects the precision of the observer's capabilities.

\begin{definition} \label{con} {\bf (Ordering on Observation Functions) }
Given two observation functions  $ {\mathcal  O}_1,  {\mathcal  O}_2$ we say that 
observation function $ {\mathcal  O}_2$ is stronger than  $ {\mathcal  O}_1$, denoted as 
 $ {\mathcal  O}_1 \preceq  {\mathcal  O}_2$ iff whenever $w,w' \in Actt^*$ such that 
  $ {\mathcal  O}_2(w) =  {\mathcal  O}_2(w)$  then also 
  $ {\mathcal  O}_1(w) =  {\mathcal  O}_1(w)$. 
\end{definition}

An ordering of observation functions characterizes the varying observational powers of an attacker, which directly corresponds to a spectrum of distinct security properties. By formalizing this hierarchy, we can systematically compare how different levels of information disclosure impact a system's vulnerability to opacity violations.

\begin{proposition}
Let $ {\mathcal  O}_1, {\mathcal  O}_2$ be two observation functions and such that $ {\mathcal  O}_2 \preceq  {\mathcal  O}_1$. Let $\phi_1, \phi_2$
be two predicates over sequences such that $\phi_2 => \phi_1$. Then
$LOp^{\phi_1}_{\mathcal O_1} \subseteq LOp^{\phi_2}_{\mathcal O_2}$.
\end{proposition}

\begin{proof}
Let $P \in LOp^{\phi_1}_{\mathcal O_1}$ and let 
 $w \in Tr(P)$ such that $\phi_2(w)$ 
holds. Since  $\phi_2 => \phi_1$ we have $\phi_1(w)$ 
holds. Since $P \in LOp^{\phi_1}_{\mathcal O_1}$ we have that 
there exists a sequence
 $w', w' \in Tr(P)$ such that $\neg \phi_1(w') $ holds
 and ${\mathcal  O_1}(w) = {\mathcal  O_1}(w')$.
 Since $\neg \phi_1  => \neg \phi_2$ we have that also 
$\neg \phi_2(w') $ holds and since   ${\mathcal  O}_2 \preceq  {\mathcal  O}_1$ we have that 
${\mathcal  O_2}(w) = {\mathcal  O_2}(w')$ and hence 
$P \in LOp^{\phi_2}_{\mathcal O_2}$.
\end{proof}

%% file: supervisor.tex
\section{Active Supervisory Control}

In this section, we will deal with the problem of what to do if a process is not secure in the sense of language opacity. Changing its design is often not a solution because of the high price, already existing hardware implementation, etc.

We will work with the concept of an active supervisor who monitors the process and in case there is a threat of leakage of sensitive information, its activity will either be completely terminated or modified so that leakage is not possible despite its continuation. In this case, it tries to insert time actions and thus precedes especially, for example, time attacks that uses time information. In most cases, our formalism allows the insertion of time actions (see rules A1, A2, Pa, S in Section 2), so that the resulting sequence of actions is still possible and therefore not suspect.
Similar to the attacker, whose observation skills are expressed by his observational function, we will also assume that the supervisor has only partial information about the system's activity, expressed by its own observational function.

\subsection{Safe traces and supervisors}

Now we define a set of safe traces of a process, which are those traces which cannot leak validity of $\phi$ under a given observation. The task of the supervisor is to ensure that the traces of process $P$ always belong to this set.

\begin{definition} \label{LK} {\bf (Safe Traces) }
Given process $P$, a predicate $\phi$ over $Actt^\star$ 
and the observation function ${\mathcal  O}$. We define 
$K^{P, \phi, {\mathcal  O}} \subseteq Tr(P)$ as $w \in K^{P, \phi, {\mathcal  O}}$ iff $\neg \phi(w)$ holds or 
there exists a sequence
 $w', w' \in Tr(P)$ such that $\neg \phi(w') $ holds
 and ${\mathcal  O}(w) = {\mathcal  O}(w')$.
\end{definition}

Clearly, we have $K^{P, \phi, {\mathcal  O}} = Tr(P)$ iff $P \in LOp^{\phi}_{\mathcal O}$, hence
if $P \not \in LOp^{\phi}_{\mathcal O}$ we have $K^{P, \phi, {\mathcal  O}} \subset Tr(P)$.
Thus, to maintain the security of a process $P$ with respect to language opacity, we must prohibit it from executing any trace in $Tr(P) \setminus K^{P, \phi, {\mathcal O}}$. To achieve this, we exploit the concept of an active supervisor $Sup$. In addition to restricting actions from a given set of controllable actions  $C$ like a conventional supervisor, an active supervisor can also insert a sequence of actions $t$ to ensure that the resulting trace performed by the supervised process always belongs to $K^{P, \phi, {\mathcal O}}$.

\begin{example}
Let $\phi(w)$ holds iff $w$ contains action $h$ and ${\mathcal  O}= {\mathcal  O}_s$ are static observation function such that  ${\mathcal  O}(h) = \epsilon$ and ${\mathcal  O}(x) = x$ for $x \not = h$. Then for $P_1= h.l.Nil$ we have $P_1 \not \in LOp^{\phi}_{\mathcal O}$ and a supervisor has to prohibit action $l$ to keep process $P_1$ secure. For  $P_2= h.l.Nil + l.Nil$ we have 
$P_2  \in LOp^{\phi}_{\mathcal O}$ and a supervisor need to do nothing. 
For  $P_3= h.l.Nil + t.l.Nil$ the supervisor has to insert action $t$ before performing $l$.
\end{example}

Similar to the attacker, we assume that the supervisor cannot observe all 
process actions. This limitation is captured by the supervisor's own, 
possibly distinct, observation function $\mathcal{O}_S$ (see 
Definition~\ref{obs}); thus, in general, $\mathcal{O}_S$ and $\mathcal{O}$ 
may be two different functions. After observing a sequence of actions via 
$\mathcal{O}_S$, the supervisor decides whether to intervene by either 
prohibiting an action or inserting a sequence of actions. If it prohibits 
an action, the process execution terminates.
We restrict the ability of the supervisor to disable only actions from  a set of controllable actions $C \subseteq A$.

We can view the supervisor $Sup_C$ (written simply as $Sup$ if $C$ is 
clear from the context) as a mapping that takes an observed trace~$w$ 
of a process~$P$ and produces another trace by inserting time delays 
or interrupting it when a controllable action occurs.
Formally, 
 $Sup_C \circ  {\mathcal  O}_S: Tr(P) \rightarrow K^{P, \phi, {\mathcal  O}}$
such that for  ${\mathcal  O}_S(w)=  x_1.x_2 \dots x_k$ we have 
$Sup(x_1.x_2 \dots x_k) = 
t^{i_1}.x_1.t^{i_2} \dots t^{i_k}.x_l.t^{i_{l+1}} $. If $l < k$ then 
$x_{l+1} \in C$,
i.e., the supervisor can stop execution only immediately before a controllable action $x_{i+1}$.

\begin{definition} {\bf (Supervisor) }
Given process $P$, a predicate $\phi$ over $Actt^\star$ 
and the observation functions ${\mathcal  O},{\mathcal  O}_S $. We say 
that $Sup_C$ is a supervisor for corresponding language opacity if 
 $Sup_C \circ  {\mathcal  O}_S(w) \in K^{P, \phi, {\mathcal  O}}$ for every $w \in Tr(P)$.
\end{definition}

Now we define a set of all supervisor which gurantee language based security for a given process $P$
and corresponding observation functions and the predicate.

\begin{definition} \label{ss} {\bf (Set of Supervisors) }
Given process $P$, a predicate $\phi$ over $Actt^\star$, $C \subset A$
and the observation functions ${\mathcal  O}, {\mathcal  O}_S$. We define 
$Sup(P, LOp^{\phi}_{\cal O}, C, {\mathcal  O}_S)$ as a set of all all corresponding supervisors.
\end{definition}

Clearly,
$Sup(P, LOp^{\phi}_{\cal O}, C, {\mathcal  O}_S) = Sup(P', LOp^{\phi}_{\cal O}, C, {\mathcal  O}_S)$  if  $P$ and   P$'$ are bisimilar, i.e. $P \sim P'$. 
The capability of a supervisor to guarantee language opacity for a given 
process $P$ depends strongly on the set $C$ of controllable actions. 
When the set $C$ is highly restrictive, the supervisor possesses fewer 
degrees of freedom to  disable transitions, what would potentially allow 
secret behaviors to be leaked to an intruder. Conversely, expanding the 
domain of controllable actions expands the supervisor's synthesis space, 
enabling the enforcement of stricter opacity constraints at the cost of 
reduced system permissiveness.

\begin{definition} \label{con} {\bf (Controllability) }
Given process $P$, a predicate $\phi$ over processes, $C \subset A$
and the observation functions ${\mathcal  O}, {\mathcal  O}_S$. We say that the
set $K^{P, \phi, {\mathcal  O}}$ is controllable iff
$w \in K^{P, \phi, {\mathcal  O}}$ and $x \in A\setminus C $ then 
$w.x \in K^{P, \phi, {\mathcal  O}}$. 
\end{definition}

Here, we present a fundamental obstacle that arises when a supervisor is structurally incapable of guaranteeing language-based opacity. This limitation typically manifests as a critical conflict between three competing constraints: the partial observability of the execution path, the restricted set of controllable actions, and the system's structural non-determinism. Specifically, if a secret execution path and a non-secret execution path produce identical observation traces for the supervisor, but diverge into an uncontrollable state for the intruder, the supervisor is left with an unsolvable dilemma. It can neither allow the actions to continue without risking immediate information leakage, nor can it disable the offending transitions without inadvertently halting benign, vital system operations. We formalize this bound on supervisory synthesis below.

\begin{proposition}
Given process $P$, a predicate $\phi$ over processes, $C \subset A$
and the observation functions ${\mathcal  O}, {\mathcal  O}_S$. Let
set $K^{P, \phi, {\mathcal  O}}$ is not controllable. Then 
$Sup(P, LOp^{\phi}_{\cal O}, C, {\mathcal  O}_S) = \emptyset$.
\end{proposition}

\begin{proof}
 Let
set $K^{P, \phi, {\mathcal  O}}$ is not controllable. Hence there exists 
$w \in K^{P, \phi, {\mathcal  O}}$ and $x \in A\setminus C $ such that
$w.x \in K^{P, \phi, {\mathcal  O}}$ i.e. no supervisor can prohibit process $P$
to perform unsafe trace.
\end{proof}

For a supervisor, it is straightforward to halt the execution of a process $P$ 
if a controllable action could lead to an unsafe trace. However, this is only 
possible if the supervisor can detect that the action is about to be performed, 
which inherently depends on its observation function $\mathcal{O}_S$. For 
instance, if $\mathcal{O}_S$ is $C$-independent (see Definition~\ref{inde}), 
the supervisor cannot directly observe when a controllable action is going to 
be executed. To determine whether a supervisor cannot  indirectly detect an action 
from $C$, we can utilize language opacity. Let $\phi(w)$ hold if and only if 
$w$ contains an action from $C$. Then, the supervisor cannot detect the occurrence 
of an action from $C$ if $P \in LOp^{\phi}_{\mathcal{O}_S}$.

On the other hand, if a supervisor can prohibit any action, then it is clearly 
a valid supervisor for any $P, \phi, \mathcal{O}, \mathcal{O}_S$. In the 
worst-case scenario, this could simply be a trivial supervisor that stops 
the execution of all traces.

\bigskip
 
\noindent {\bf Corollary} Given process $P$, a predicate $\phi$ over processes, $C = A$
and the observation functions ${\mathcal  O}, {\mathcal  O}_S$. Then 
$Sup(P,LPOp^{\phi}_{\cal O}, C, {\mathcal  O}_S) \not = \emptyset$.

\bigskip
 
To guarantee a minimal restriction of process behavior our 
 aim is to find a maximal process supervisor in the sense that it minimally restricts the behavior of the original process. The formal definition is the following.

\begin{definition} \label{mpso} {\bf (Supervisors Order and Maximal Supervisor)} 
Let $Sup_1, Sup_2 \in Sup(P, LOp^{\phi}_{\cal O}, C, {\mathcal  O}_S) $. 
We say that supervisor $Sup_1$ is more restrictive (denoted by 
$Sup_1 \ll Sup_2$
then $Sup_2$
if 
$Tr( {\mathcal  C}(P,{\mathcal  O_S}, Sup_1)) \subseteq
Tr( {\mathcal  C}(P,{\mathcal  O_S}, Sup_2)$.
Process $ Sup \in Sup(P, LOp^{\phi}_{\cal O}, C, {\mathcal  O}_S) $ 
is called maximal process supervisor for language  opacity $LOp^{\phi}_{\cal O}$ 
iff for every $ Sup' \in Sup(P, LOp^{\phi}_{\cal O}, C, {\mathcal  O}_S) $
we have 
$Tr( {\mathcal  C}(P,{\mathcal  O_S}, Sup')) \subseteq Tr( {\mathcal  C}(P,{\mathcal  O_S}, Sup)) $. 
\end{definition}


Unfortunately it is undecidable to verify whether the process $Sup$ is a 
 process 
supervisors for $P$ and language opacity   as it is stated by the following proposition.

\begin{proposition} \label{undec}
The property that $Sup$ is a process supervisor for language opacity for process $P$ is undecidable in general.
\end{proposition}

 \begin{proof}
The proof is based on the idea that already language opacity is undecidable (see Proposition 2. in \cite{Gru15a}). Suppose that the property is decidable.
Let $Sup = \mu X . \sum_{x \in Actt} x'.x.X$ i.e $Sup$ does not restrict anything. We have that $Sup$ is a 
  supervisor for language opacity for process $P$ iff 
$ P \in LOp^{\phi}_{\cal O})$. Hence we would be able to decide language opacity what contradicts its undecidability. 
\end{proof}

By a similar argument, we can prove the following statement, which claims that we cannot even decide whether there is at least one supervision that guarantees the security of systems.

\begin{proposition} \label{undecs}
It is undecidable whether 
$Sup(P, LOp^{\phi}_{\cal O}, C, {\mathcal  O}_S)  = \emptyset$ in general.
\end{proposition}

In the proposition that follows, we articulate the underlying assumptions that rigorously guarantee the existence of a nontrivial supervisor. 

\begin{proposition}
Given process $P$, a predicate $\phi$ over processes
and the observation functions ${\mathcal  O}, {\mathcal  O}_S$ such that ${\mathcal  O} \preceq {\mathcal  O}_S$ and $C=A$. Then \linebreak 
$Sup(P, LOp^{\phi}_{\cal O}, C, {\mathcal  O}_S)$ contains at least one nontrivial supervisor, i.e. such which does not prohibit all actions. 
\end{proposition}

\begin{proof}
Let $w \in \mathit{Tr}(P) \setminus K^{P, \phi, \mathcal{O}}$. If no such trace $w$ exists, then a  supervisor that neither prohibits any actions nor inserts any sequences trivially guarantees the security of $P$. Otherwise, the objective of the supervisor is to prevent $P$ from executing any such leakage trace $w$. 

Since $\mathcal{O} \preceq \mathcal{O}_S$, the supervisor possesses at least as much observational precision as the attacker. Consequently, the supervisor can precisely identify and disable only those controllable actions that would otherwise allow an intruder to deduce the validity of the secret predicate $\phi$.
\end{proof}

\subsection{Observations, Predicates and Supervisor as Processes}

In this subsection, we formalize observation functions, predicates, and 
supervisors by representing them as processes. Owing to the Turing-complete 
expressive power of Timed Process Algebra (TPA) processes, this framework 
allows us to explicitly model any computable observation function, predicate, 
or supervisor. 

These components are conceptualized as reactive processes that interact 
with the underlying system execution. Specifically, they take a system trace 
of $P$ as input and process it according to their designated roles:
\begin{itemize}
    \item \textbf{Observation functions} output the corresponding observable 
    counterpart of the trace;
    \item \textbf{Predicates} emit a distinct, dedicated action to signal 
    that the current trace satisfies the secret property $\phi$;
    \item \textbf{Supervisors} dynamically monitor the execution and output 
    a modified or restricted trace to enforce safety and opacity requirements.
\end{itemize}

By shifting these concepts 
into a uniform, process-algebraic framework, we can leverage standard 
verification techniques to analyze their interactions. Based on this unified 
representation, we then establish several decidability results concerning 
both the verification of language opacity and the synthesis of valid 
supervisors (see Fig. \ref{monitor}).

\begin{figure}[htp]
\centering {\tt    \setlength{\unitlength}{1,3pt}
\begin{picture}(220,30)

\put(45,10){\framebox(25,20){${\mathcal  O_S}$}}

\put(75,10){\framebox(25,20){$Sup$}}

\put(0,16){\vector(1,0){15}}

\put(15,24){\vector(-1,0){15}}



\put(15,5){\framebox(90,30){$ $}}

\put(40,16){\line(1,0){5}}

\put(40,24){\line(1,0){5}}

\put(70,24){\line(1,0){5}}
\put(70,16){\line(1,0){5}}

\put(20,10){\framebox(20,20){$P$}}


\put(120,16){\vector(1,0){15}}

\put(135,24){\vector(-1,0){15}}


\put(135,5){\framebox(60,30){$ P$}}




\end{picture}}
\caption{Supervisory Control} \label{monitor}
\end{figure}

To orchestrate the behavior of these concurrent entities, we introduce the 
notion of (process) contexts. Contexts serve as the formal environment that 
encapsulates and governs the communication interfaces between the core system 
process, the observation function, predicates, and the supervisor. 

By utilizing parallel composition alongside appropriate restriction and 
relabeling operators, a context strictly regulates how actions are intercepted, 
transformed, and observed. For instance, it ensures that the execution traces 
generated by the process $P$ are systematically routed as inputs to both the 
observation function and the supervisor, while simultaneously restricting 
uncoordinated or illegal actions. This algebraic encapsulation allows us to 
model the closed-loop system behavior precisely, providing a clean foundation 
for analyzing properties such as opacity and supervisor synthesis.
By context $ {\mathcal  C}$ we mean a process term with  placeholders ${\cal
H}$. Formally, the set of TPA simple contexts is defined by the following BNF notation:

$$ {\mathcal  C} \, \, ::= \, \,   {\cal H}_i  \, \,  |  \, \, op( {\mathcal  C}_1, {\mathcal  C}_2, \dots {\mathcal  C}_n)  \, \,
 $$

\noindent where  
  ${\mathcal  C}, {\mathcal  C}_1, \dots  {\mathcal  C}_n$ are  TPA contexts,  $op \in \{ 
[S], \setminus, |\}$ (i.e. operations relabelling, restriction and parallel composition)
and ${\cal H}$ is the place holder. By
$ {\mathcal  C}(P)$ we denote process obtained from process simple context ${\mathcal  C}$ and
process $P$ by  substituting $P$ by place holders ${\cal H}_1, {\cal H}_2$, ${\cal H}_3$ , i.e.
${\mathcal  C}(P, O,  Sup) = {\mathcal  C}[P/{\cal H}_1, O/{\cal H}_2, Sup/{\cal H}_3 ]$.
We require that $Tr({\mathcal C}(P,{\mathcal O}_S, Sup)) \subseteq K^{P, \phi, {\mathcal O}} Tr(P)$, 
i.e., that the supervised process $P$ performs only safe traces. 
Later, we will investigate the existence of such a supervisor $Sup$.

\begin{proposition}
Let $Sup$ and $Sup'$ be two  supervisors, and let $Sup \sim Sup'$.
Then 
$Tr( {\mathcal  C}(P,{\mathcal  O_S}, Sup)) = Tr( {\mathcal  C}(P,{\mathcal  O_S},Sup'))$.

\end{proposition}

\begin{proof}
  The proof follows from the fact that bisimulation $\sim$ is a congruence 
relation and is stronger than trace equivalence.
\end{proof}

\noindent {\bf Corollary} For any supervisor $Sup$, if $Sup \sim Q$ 
then also $Q$ is a supervisor.

\bigskip

To obtain a decidable variant of the previous propositions, we first place 
some restrictions on trace predicates. We model predicates by special 
processes called tests. These tests communicate with the process's trace 
and produce a $\surd$ action if the corresponding predicates hold for 
the trace. In the subsequent proposition, we show how to exploit this 
idea for process opacity.

\begin{definition}
We say that the process $T_\phi$ is the test representing the predicate
$\phi$ if $\phi(w)$ holds iff $(w.Nil | T_\phi) \setminus At \approx_{w}
\surd.Nil$ where $\surd$ is a new action indicating a passing of the
test. If $T_\phi$ is the finite state process we say that $\phi$ is
the finitely definable predicate.
\end{definition}

Suppose that both $\phi$ and $\neg \phi$ are finitely definable 
predicates. Then, under some additional conditions, we can reduce the 
tasks of checking whether $Sup$ is a supervisor, comparing two 
supervisors, or determining if a supervisor is maximal to the problem 
of checking trace inclusion, which is decidable for finite automata 
(see \cite{TLSG18}). We obtain the following results.
Moreover if $Sup_1 \prec_s Sup_2$ (see Definition \ref{simulation}) we have 
$Sup_1 \ll Sup_2$.

\begin{proposition}
Let $\phi$ and $\neg \phi$ be finitely definable predicates, and let 
${\cal O}$ and ${\cal O}_S$ be static. The property that $Sup$ is a 
supervisor for language opacity for a finite-state process $P$ is 
decidable. The ordering $\ll$ of supervisors, as well as the property 
of being a maximal supervisor, is also decidable.
\end{proposition}

Moreover, for  static  observation functions ${\cal O}$, ${\cal O}_S$
and  $\phi$ and $\neg \phi$  finitely definable predicates there exists
 finite state maximal process supervisor for language opacity for any finite state process $P$.  This follows from the fact that such observation function can be emulated
 by finite-state processes since only finite memory is required.

\begin{proposition} \label{maxs}
Let $\phi$ and $\neg \phi$ are finitely definable predicates and ${\cal O}$, ${\cal O}_S$
are static.
Then for any finite-state process $P$  there exists finite-state process $Sup$ 
which is the maximal supervisor for corresponding language opacity. 
\end{proposition}

Note that the above-mentioned properties can be directly extended to $m$-orwellian observation functions, where the observer's current visibility is boundedly dependent on a finite history of the past $m$ actions. However, transitioning to general dynamic and orwellian observation functions introduces severe computational challenges. In these settings, the observation mapping can change dynamically based on the entire execution history, or even retroactively based on future actions (in the case of orwellian functions). Consequently, even if these observation functions are assumed to be computable, deciding security properties like opacity or non-interference under their scope typically requires the full expressiveness of \emph{Timed-Place Automata} (TPA) or tracking infinite state spaces. As a result, the corresponding verification and synthesis properties remain undecidable in the general case, necessitating either structural restrictions on the system or the use of bounded-horizon approximations.

It is well known that both trace equivalence and bisimulation are 
congruences (see \cite{Mil89}), meaning we can replace any component in 
the process ${\mathcal C}(P,{\mathcal O}_S, Sup)$ with an equivalent one 
without changing its functionality; i.e., 
$Tr({\mathcal C}(P,{\mathcal O}'_S, Sup')) \subseteq K^{P, \phi, {\mathcal O}} Tr(P)$ 
will still hold.

Regarding the persistent variant of language-based opacity (see Definition \ref{def:persistent_opacity}), the enforcement mechanism requires a fundamentally different formulation of the set of safe traces. Specifically, a system trace is defined as safe if and only if it satisfies a dual condition: first, its execution must not leak the validity of the secret predicate $\phi$ to an adversary, and second, the subsequent state or residual process reached after executing this trace must itself remain language opaque. Computing this restricted set of safe traces is inherently computationally more demanding than in the non-persistent case, as it necessitates a recursive or fixed-point analysis of the process's future behavior. Nevertheless, under the structural and environmental restrictions outlined above, the synthesis of this set remains theoretically and practically feasible.

\subsection{Timing Attacks}

Our formal model can also account for the time-dependent aspects of system behavior, process opacity can also be used to express vulnerabilities to timing attacks.
These attacks use 
timing information leakage, which is the ability of an attacker to deduce internal (private) information depending on timing information.
They, as side-channel attacks, represent a serious threat to many systems.
They allow intruders 
 ``break'' ``unbreakable'' systems, 
algorithms, protocols, etc. 
For example, by carefully measuring the amount of time required to
perform private key operations, 
attackers may be able to find fixed Diffie-Hellman exponents, factor
RSA keys, and break other  
cryptosystems (see \cite{Ko96}).
This idea was developed in \cite{DKL98} where a timing attack 
against smart card implementation of RSA was conducted.
In \cite{HH99}, a timing attack on the RC5 block encryption algorithm
is described. The analysis is motivated by 
the possibility that some implementations of RC5 could result in 
data-dependent rotations taking a 
 time that is a function of the data.
In \cite{HK99},  the vulnerability of two implementations of the Data Encryption Standard 
(DES) cryptosystem under a timing attack is studied.
It is shown that a timing attack yields the Hamming weight of the key used by both 
DES implementations. Moreover, the attack is computationally inexpensive. 
A timing attack against an implementation
of AES candidate Rijndael is described
in \cite{KQ99}, and the one against the popular SSH 
protocol in \cite{SWT01}. 
In \cite{BM06}
 several novel timing attacks against the common table-driven software implementation of the AES cipher are described. 
Also  possible attacks on most of the currently used processors (Meltdown and Spectre)  belong to timing attacks.
Timing attacks on web privacy and some corresponding formal models can be found in \cite{FGL02}.
 Let
${\cal O}$, ${\cal O'}$  be observation functions such that 
${\cal O}(w|_{Act})={\cal O'}(w)$ where  $w|_{Act}$ 
represents the sequence $w$ without $t$ actions.
Then $P$ is prone to timing  attacks iff 
$P \not \in LOp^{\phi}_{\mathcal O}$ but $P  \in LOp^{\phi}_{\mathcal O'}$. 
Consequently, the elapsed time becomes a side channel, where the duration of the computation itself reveals information about the validity of $\phi$. For instance, a shorter execution time for a particular input might indicate that a certain condition related to $\phi$ was met (or not met), allowing the attacker to deduce properties of the secret."


Let $w,w' \in Actt^*$ such  that we can obtain $w$ from $w'$ by removing som t actions from $w'$. We say that $w$ is time subsequence of $W$, denoted $w \prec_t w'$. For example, we have  $atbc \prec_t tattbtct$.
We say that predicate $\phi$ over sequences from $Actt^*$ is time dependable if for every $w \in Actt^*$ such that $\phi(w)$ holds
then there exists $w', w \prec_t w'$ such that $\neg \phi(w')$ holds.

\begin{proposition}
Let $P$ is prone to timming attack with respect to ${\mathcal O}$
and time dependable predicate $\phi$. Let ${\mathcal O} \preceq {\mathcal O_S}$. Then there exist a supervisor for $P$ with empty controlable set $C$.
\end{proposition}

\begin{proof}
Let for every $w \in Tr(P) \setminus K^{P, \phi, {\mathcal  O}}$ there exists $w', w \prec_t w'$ such that $\neg \phi(w')$ holds. Clearly, $w' \in Tr(P)$ (see transition rules A1, A2, Pa, S) and hence it is enough to insert corresponding $t$ actions.  
\end{proof}

%% file: conclusion.tex
\section{Conclusions}


In this paper, we have investigated the framework of active time-inserting supervisors within a timed process algebra setting, specifically designed to enforce language-based security properties and mitigate timing attacks. By possessing the dual capability to selectively disable actions and dynamically introduce strategic time delays into the system's execution stream, these supervisors offer a powerful mechanism for neutralizing timing side-channels. We have formally characterized the conditions governing the existence of such time-inserting supervisors for a given process, under specified supervisor and attacker observation functions, and with respect to a target security predicate over process traces.

Moving forward, several promising avenues for future research emerge. A primary objective is the formulation and synthesis of \emph{minimal} time-inserting supervisors—that is, supervisors that inject the absolute minimum temporal overhead necessary to robustly guarantee language-based opacity. Minimizing these delays is critical to maintaining system performance and QoS (Quality of Service). To achieve this optimization, we plan to adapt and extend algorithmic resource-minimization techniques similar to those proposed in \cite{JIL19}. 

Furthermore, the compositional nature of our process algebraic approach opens up possibilities for integrating other multi-dimensional parameters. By enriching the underlying algebra with operators that capture spatial distribution, network architectures, or power consumption metrics, this framework can be extended to verify multi-domain security properties. Investigating the trade-offs between security enforcement, temporal delays, and energy expenditure will allow us to derive formal models that possess not only rigorous theoretical foundations but also profound practical utility for securing modern embedded and cyber-physical systems.

The theoretical framework established in this paper provides a rigorous foundation 
for analyzing system security within a process-algebraic context. By explicitly 
delineating the boundaries of feasible system protection, our work characterizes 
the exact conditions under which security properties can be guaranteed. A key 
contribution lies in the simultaneous modeling of both attackers and supervisors 
equipped with powerful, asymmetric observation functions. This dual perspective 
captures the realistic and adversarial dynamics of modern decentralized systems, 
where both defensive and offensive agents operate under partial visibility.

For a given process,  observation functions, and a target security 
predicate, the verification of the studied properties can be automated using 
appropriate software verification tools. For instance, when observation 
functions are themselves modeled as processes and the predicate $\phi$ is 
expressed via $\mu$-calculus formulae \cite{Sti96}, the CAAL (Concurrency 
Workbench Alternating-Time Logic) system \cite{CAAL} provides a powerful 
environment for this purpose. Utilizing CAAL's integrated model checker, 
alongside its capabilities for checking trace equivalence and strong/weak 
bisimulation, allows us to formally validate the system's behavior against 
the defined security requirements. 

Consequently, this framework enables the exact verification of the proposed 
supervisor's properties as well. By modeling the composition of the plant, 
the supervisor, and the attacker within the tool, we can systematically 
check whether the synthesized supervisor effectively enforces the predicate 
$\phi$ under the given observation constraints, while ensuring that the 
controlled system remains  compliant with the specified 
security goals.

The theoretical insights obtained in this study offer meaningful guidance for 
the synthesis of practical supervisors. Specifically, our results clarify the 
fundamental trade-offs between supervisory control and system utility, illustrating 
what can be realistically expected from an enforcement mechanism under specific 
information constraints. Furthermore, by uncovering the underlying mathematical 
principles behind these capabilities, this work bridges the gap between abstract 
process security models and the design of robust, resilient real-world architectures. 
Future work will focus on scaling these verification techniques and exploring the 
algorithmic complexity of supervisor synthesis under dynamic observation policies.

\bibliography{IS.bib}
\bibliographystyle{plain}

%% file: IS.bib
@article{RRF21,
  doi = {10.48550/ARXIV.2102.09338},
  
  url = {https://arxiv.org/abs/2102.09338},
  
  author = {Rashidinejad, Aida and Reniers, Michel and Fabian, Martin},
  
  title = {Supervisory Control Synthesis of Timed Automata Using Forcible Events},
  
  publisher = {arXiv},
  
  year = {2021},
  
  copyright = {arXiv.org perpetual, non-exclusive license}
}

@article{BKR04,
title = {Modelling Opacity Using Petri Nets},
journal = {Electronic Notes in Theoretical Computer Science},
volume = {121},
pages = {101-115},
year = {2005},
note = {Proceedings of the 2nd International Workshop on Security Issues with Petri Nets and other Computational Models (WISP 2004)},
issn = {1571-0661},
doi = {https://doi.org/10.1016/j.entcs.2004.10.010},
url = {https://www.sciencedirect.com/science/article/pii/S1571066105000277},
author = {Jeremy W. Bryans and Maciej Koutny and Peter Y.A. Ryan}
}

@inproceedings{BKMR06,
author = {Bryans, Jeremy and Koutny, Maciej and Mazare, Laurent and Ryan, Peter},
year = {2008},
month = {11},
pages = {421-435},
title = {Opacity Generalised to Transition Systems},
volume = {7},
isbn = {978-3-540-32628-1},
journal = {Int. J. Inf. Sec.},
doi = {10.1007/11679219_7}
}

@inproceedings{DKL98,
author="Dhem, Jean-Fran{\c{c}}ois
and Koeune, Fran{\c{c}}ois
and Leroux, Philippe-Alexandre
and Mestr{\'e}, Patrick
and Quisquater, Jean-Jacques
and Willems, Jean-Louis",
editor="Quisquater, Jean-Jacques
and Schneier, Bruce",
title="A Practical Implementation of the Timing Attack",
booktitle="Smart Card Research and Applications",
year="2000",
publisher="Springer Berlin Heidelberg",
address="Berlin, Heidelberg",
pages="167--182",
isbn="978-3-540-44534-0"
}

@article{FGL02,
  author    = {Riccardo Focardi and
               Roberto Gorrieri and
               Ruggero Lanotte and
               Andrea Maggiolo{-}Schettini and
               Fabio Martinelli and
               Simone Tini and
               Enrico Tronci},
  title     = {Formal Models of Timing Attacks on Web Privacy},
  journal   = {Electron. Notes Theor. Comput. Sci.},
  volume    = {62},
  pages     = {229--243},
  year      = {2001},
  url       = {https://doi.org/10.1016/S1571-0661(04)00329-9},
  doi       = {10.1016/S1571-0661(04)00329-9},
  bibsource = {dblp computer science bibliography, https://dblp.org}
  }

@INPROCEEDINGS{fabio,
  author={Focardi, R. and Gorrieri, R. and Martinelli, F.},
  booktitle={Proceedings 13th IEEE Computer Security Foundations Workshop. CSFW-13}, 
  title={Information flow analysis in a discrete-time process algebra}, 
  year={2000},
  volume={},
  number={},
  pages={170-184},
  doi={10.1109/CSFW.2000.856935}}

@article{gor04,
author = {Gorrieri, Roberto and Martinelli, Fabio},
year = {2004},
month = {03},
pages = {23-49},
title = {A simple framework for real-time cryptographic protocol analysis with compositional proof rules},
volume = {50},
journal = {Science of Computer Programming},
doi = {10.1016/j.scico.2004.01.001}
}

@inproceedings{Gru21a,
  author    = {Damas P. Gruska},
  editor    = {Ladjel Bellatreche and
               George A. Chernishev and
               Antonio Corral and
               Samir Ouchani and
               J{\"{u}}ri Vain},
  title     = {Time Insertion Functions},
booktitle = {Advances in Model and Data Engineering in the Digitalization Era -
               {MEDI} 2021 International Workshops: DETECT, SIAS, CSMML, BIOC, HEDA,
               Tallinn, Estonia, June 21-23, 2021, Proceedings},
  series    = {Communications in Computer and Information Science},
  volume    = {1481},
  pages     = {181--188},
  publisher = {Springer},
  year      = {2021},
  url       = {https://doi.org/10.1007/978-3-030-87657-9\_14},
  doi       = {10.1007/978-3-030-87657-9\_14},
  bibsource = {dblp computer science bibliography, https://dblp.org}
}

@inproceedings{Gru19,
  author    = {Damas P. Gruska},
  editor    = {Yannis Manolopoulos and
               George Angelos Papadopoulos and
               Athena Stassopoulou and
               Ioanna Dionysiou and
               Ioannis Kyriakides and
               Nicolas Tsapatsoulis},
  title     = {Security and time insertion},
  booktitle = {Proceedings of the 23rd Pan-Hellenic Conference on Informatics, PCI
               2019, Nicosia, Cyprus, November 28-30, 2019},
  pages     = {154--157},
  publisher = {{ACM}},
  year      = {2019},
  url       = {https://doi.org/10.1145/3368640.3368668},
  doi       = {10.1145/3368640.3368668},
  bibsource = {dblp computer science bibliography, https://dblp.org}
}

@inproceedings{GR18,
  author    = {Damas P. Gruska and
               M. Carmen Ruiz},
  editor    = {Bernd{-}Holger Schlingloff and
               Samira Akili},
  title     = {Opacity-enforcing for Process Algebras},
  booktitle = {Proceedings of the 27th International Workshop on Concurrency, Specification
               and Programming, Berlin, Germany, September 24-26, 2018},
  series    = {{CEUR} Workshop Proceedings},
  volume    = {2240},
  publisher = {CEUR-WS.org},
  year      = {2018},
  url       = {http://ceur-ws.org/Vol-2240/paper1.pdf},
  bibsource = {dblp computer science bibliography, https://dblp.org}
}

@inproceedings{Gru15,
  author    = {Damas P. Gruska},
  editor    = {Andrei Voronkov and
               Irina B. Virbitskaite},
  title     = {Process Opacity for Timed Process Algebra},
  booktitle = {Perspectives of System Informatics - 9th International Ershov Informatics
               Conference, {PSI} 2014, St. Petersburg, Russia, June 24-27, 2014.
               Revised Selected Papers},
  series    = {Lecture Notes in Computer Science},
  volume    = {8974},
  pages     = {151--160},
  publisher = {Springer},
  year      = {2014},
  url       = {https://doi.org/10.1007/978-3-662-46823-4\_13},
  doi       = {10.1007/978-3-662-46823-4\_13},
  bibsource = {dblp computer science bibliography, https://dblp.org}
  }

@inproceedings{Gru15a,
  author    = {Damas P. Gruska},
  editor    = {Manuel Mazzara and
               Andrei Voronkov},
  title     = {Dynamics Security Policies and Process Opacity for Timed Process Algebras},
  booktitle = {Perspectives of System Informatics - 10th International Andrei Ershov
               Informatics Conference, {PSI} 2015, in Memory of Helmut Veith, Kazan
               and Innopolis, Russia, August 24-27, 2015, Revised Selected Papers},
  series    = {Lecture Notes in Computer Science},
  volume    = {9609},
  pages     = {149--157},
  publisher = {Springer},
  year      = {2015},
  url       = {https://doi.org/10.1007/978-3-319-41579-6\_12},
  doi       = {10.1007/978-3-319-41579-6\_12},
  bibsource = {dblp computer science bibliography, https://dblp.org}
}

@inproceedings{HH99,
author = {Handschuh, Helena and Heys, Howard M.},
title = {A Timing Attack on RC5},
year = {1998},
isbn = {3540658947},
publisher = {Springer-Verlag},
address = {Berlin, Heidelberg},
booktitle = {Proceedings of the Selected Areas in Cryptography},
pages = {306–318},
numpages = {13},
series = {SAC '98}
}

@article{HK99,
author = {Hevia, Alejandro and Kiwi, Marcos},
title = {Strength of Two Data Encryption Standard Implementations under Timing Attacks},
year = {1999},
issue_date = {Nov. 1999},
publisher = {Association for Computing Machinery},
address = {New York, NY, USA},
volume = {2},
number = {4},
issn = {1094-9224},
url = {https://doi.org/10.1145/330382.330390},
doi = {10.1145/330382.330390},
journal = {ACM Trans. Inf. Syst. Secur.},
month = nov,
pages = {416-437},
numpages = {22}
}

@article{JIL19,
title = {Enforcing opacity by insertion functions under multiple energy constraints},
journal = {Automatica},
volume = {108},
pages = {108476},
year = {2019},
issn = {0005-1098},
doi = {https://doi.org/10.1016/j.automatica.2019.06.028},
url = {https://www.sciencedirect.com/science/article/pii/S0005109819303243},
author = {Yiding Ji and Xiang Yin and St{\'{e}}phane Lafortune}
}

@InProceedings{Ko96,
author="Kocher, Paul C.",
editor="Koblitz, Neal",
title="Timing Attacks on Implementations of Diffie-Hellman, RSA, DSS, and Other Systems",
booktitle="Advances in Cryptology --- CRYPTO '96",
year="1996",
publisher="Springer Berlin Heidelberg",
address="Berlin, Heidelberg",
pages="104--113",
isbn="978-3-540-68697-2"
}

@book{Mil89,
author = {Milner, R.},
title = {Communication and Concurrency},
year = {1989},
isbn = {0131149849},
publisher = {Prentice-Hall, Inc.},
address = {USA}
}

@ARTICLE{RW89,
  author={Ramadge, P.J.G. and Wonham, W.M.},
  journal={Proceedings of the IEEE}, 
  title={The control of discrete event systems}, 
  year={1989},
  volume={77},
  number={1},
  pages={81-98},
  doi={10.1109/5.21072}}

@inproceedings{SWT01,
author = {Song, Dawn Xiaodong and Wagner, David and Tian, Xuqing},
title = {Timing Analysis of Keystrokes and Timing Attacks on SSH},
year = {2001},
publisher = {USENIX Association},
address = {USA},
booktitle = {Proceedings of the 10th Conference on USENIX Security Symposium - Volume 10},
articleno = {25},
location = {Washington, D.C.},
series = {SSYM'01}
}

@TECHREPORT{KQ99,
    author = {Francois Koeune and Francois Koeune and Jean-Jacques Quisquater and Jean-jacques Quisquater},
    title = {A timing attack against Rijndael},
    institution = {},
    year = {1999}
}

@article{TLSG18,
  author    = {Yin Tong and
               Zhiwu Li and
               Carla Seatzu and
               Alessandro Giua},
  title     = {Current-state opacity enforcement in discrete event systems under
               incomparable observations},
  journal   = {Discret. Event Dyn. Syst.},
  volume    = {28},
  number    = {2},
  pages     = {161--182},
  year      = {2018},
  url       = {https://doi.org/10.1007/s10626-017-0264-7},
  doi       = {10.1007/s10626-017-0264-7},
  bibsource = {dblp computer science bibliography, https://dblp.org}
}

@inbook{W70,
author = {Wonham, Walter},
year = {1970},
month = {01},
pages = {542-562},
title = {On the control of discrete-event systems},
volume = {135},
isbn = {3-540-51605-0},
doi = {10.1007/BFb0008476}
}

@article{KC24,
  title={Marking Data-Informativity and Data-Driven Supervisory Control of Discrete-Event Systems},
  author={Kuma Fuchiwaki and Kai Cai},
  journal={Proceedings of the Japan Joint Automatic Control Conference},
  volume={67},
  number={ },
  pages={102-107},
  year={2024},
  doi={10.11511/jacc.67.0_102}
}

@ARTICLE{XTWS25,
  author={Xie, Gang and Tong, Yin and Wang, Xiaomin and Seatzu, Carla},
  journal={IEEE Transactions on Automatic Control}, 
  title={Resilient Supervisor Synthesis for Labeled Petri Nets Against Sensor Attacks}, 
  year={2025},
  volume={},
  number={},
  pages={1-8},
}

@article{HAM15,
title = {Enforcing Diagnosability in Interpreted Petri Nets},
journal = {IFAC-PapersOnLine},
volume = {48},
number = {7},
pages = {58-63},
year = {2015},
note = {5th IFAC International Workshop on Dependable Control of Discrete Systems},
issn = {2405-8963},
doi = {https://doi.org/10.1016/j.ifacol.2015.06.473},
url = {https://www.sciencedirect.com/science/article/pii/S2405896315007120},
author = {K. Hernández-Rueda and M.E. Meda-Campaña and J. Arámburo-Lizárraga},
}

@INPROCEEDINGS{GM82,
  author={Goguen, J. A. and Meseguer, J.},
  booktitle={1982 IEEE Symposium on Security and Privacy}, 
  title={Security Policies and Security Models}, 
  year={1982},
  volume={},
  number={},
  pages={11-11},
  doi={10.1109/SP.1982.10014}}

@InProceedings{BM06,
author="Bonneau, Joseph
and Mironov, Ilya",
editor="Goubin, Louis
and Matsui, Mitsuru",
title="Cache-Collision Timing Attacks Against AES",
booktitle="Cryptographic Hardware and Embedded Systems - CHES 2006",
year="2006",
publisher="Springer Berlin Heidelberg",
address="Berlin, Heidelberg",
pages="201--215",
isbn="978-3-540-46561-4"
}

@inproceedings{Gru26,
  author       = {Damas P. Gruska},
  editor       = {Maurice H. ter Beek and
                  Leopoldo Teixeira},
  title        = {State-Based Security and Time-Inserting Supervisors},
  booktitle    = {Formal Methods: Foundations and Applications - 28th Brazilian Symposium,
                  {SBMF} 2025, Recife, Brazil, December 3-5, 2025, Proceedings},
  series       = {Lecture Notes in Computer Science},
  volume       = {16363},
  pages        = {3--18},
  publisher    = {Springer},
  year         = {2025},
  url          = {https://doi.org/10.1007/978-3-032-12086-1\_1},
  doi          = {10.1007/978-3-032-12086-1\_1},
  bibsource    = {dblp computer science bibliography, https://dblp.org}
}

@InProceedings{CAAL,
author="Andersen, Jesper R.
and Andersen, Nicklas
and Enevoldsen, S{\o}ren
and Hansen, Mathias M.
and Larsen, Kim G.
and Olesen, Simon R.
and Srba, Jir{\'i}
and Wortmann, Jacob K.",
editor="Leucker, Martin
and Rueda, Camilo
and Valencia, Frank D.",
title="CAAL: Concurrency Workbench, Aalborg Edition",
booktitle="Theoretical Aspects of Computing - ICTAC 2015",
year="2015",
publisher="Springer International Publishing",
address="Cham",
pages="573--582",
isbn="978-3-319-25150-9"
}

@Inbook{Sti96,
author="Stirling, Colin",
editor="Moller, Faron
and Birtwistle, Graham",
title="Modal and temporal logics for processes",
bookTitle="Logics for Concurrency: Structure versus Automata",
year="1996",
publisher="Springer Berlin Heidelberg",
address="Berlin, Heidelberg",
pages="149--237",
isbn="978-3-540-49675-5",
doi="10.1007/3-540-60915-6_5",
url="https://doi.org/10.1007/3-540-60915-6_5"
}
